# Cohesive self-organization of mobile microrobotic swarms

Berk Yigit[#], Yunus Alapan[#], and Metin Sitti[*]

Physical Intelligence Department, Max Planck Institute for Intelligent Systems, 70569 Stuttgart, Germany

[*] E-mail: sitti@is.mpg.de

[#] The authors contributed equally and share the first authorship.

**Abstract**

Mobile microrobots are envisioned to be useful in a wide range of high-impact applications, many of which requiring cohesive group formation to maintain self-bounded swarms in the absence of confining boundaries. Cohesive group formation relies on a balance between attractive and repulsive interactions between agents. We found that a balance of magnetic dipolar attraction and multipolar repulsion between self-assembled particle chain microrobots enable their self-organization into cohesive clusters. Self-organized microrobotic clusters translate above a solid substrate via a hydrodynamic self-propulsion mechanism. Cluster velocity increases with cluster size, resulting from collective hydrodynamic effects. Clustering is promoted by the strength of cohesive interactions and hindered by heterogeneities of individual microrobots. Scalability of cohesive interactions allows formation of larger groups, whose internal spatiotemporal organization undergoes a transition from solid-like ordering to liquid-like behavior with increasing cluster size. Our work elucidates the dynamics of clustering under cohesive interactions, and presents an approach for addressing operation of microrobots as localized teams.

Untethered mobile microrobotic systems are envisioned to revolutionize our ability to manipulate the microscopic world with unprecedented flexibility. Mobile microrobots, actuated with external magnetic and acoustic fields, light, chemical reactions, and biological propellers, have been developed recently for precision micromanipulation, minimally invasive medical operations, and environmental applications[1-8]. However, translation of microrobots to most of these applications requires large numbers of microrobots (with sizes smaller than 50 µm) to work together to manipulate their macroscopic targets in much larger millimeter and centimeter scales.

Microrobot swarms have been introduced to address the need for collective functions and navigation of large numbers of microrobots in complex environments. Microrobot swarms of magnetic micro/nanoparticles have been utilized for enhancing functional output, and their reconfigurability improved multifunctionality and adaptability to dynamic environments[9-13]. For enhanced mobility and imaging, previous approaches relied on attractive interactions to keep the particles aggregated. However, in the absence of balancing attractive and repulsive interactions, these swarms are limited to disordered formations, lacking cohesive self-organization inherent in natural swarms[14-16]. Cohesive self-organization facilitates group formation in open spaces via attraction at large separations, and prevents jamming, overcrowding, and clumping at high densities via repulsion at small separations between agents[17].

To emulate the bio-inspired self-organized cohesion in synthetic swarms, distance-dependent interactions between microrobotic agents need to be controlled via physical forces. Here, we demonstrate self-organization of cohesive microrobotic teams emerging from their magnetic multipolar interactions in a liquid environment (**Fig. 1**). Formation, propulsion, and interactions of microrobots were controlled by a global precessing magnetic field. Microrobots were formed by the dynamic self-assembly of paramagnetic microparticles into anisotropic linear chains. The interactions between the microrobotic chains were controlled via their induced magnetic moments. At specific opening angles of the precessing field, magnetic interactions between chains carry a slowly decaying dipolar attraction and a rapidly decaying multipolar repulsion. These counteracting effects give rise to a steady-state distance between pairs, where the sum of dipolar and multipolar forces equates to zero. Under cohesive interactions, chains self-organize into clusters by arranging themselves at steady-state distances from their neighbors. These clusters leverage collective synergies to move faster as cluster size increases, which is promoted by hydrodynamic interactions. Group formation and dissolution was mainly determined by the competition between cohesive interactions and intragroup heterogeneities. Internal organization of clusters ranged from solid-like ordering to liquid-like dynamic behavior, which depended on group size. Our approach addresses the operation of microrobots as localized teams, which will inspire researchers interested in active matter and microrobotics applications for designing advanced collective systems.

**Results**

**Cohesive interactions of chain microrobots**

Microrobots were formed by the dynamic self-assembly of superparamagnetic particles into linear chains under a global precessing magnetic field. Initially, a low concentration of monodisperse superparamagnetic particles (particle radius $a$ was approximately 2.5 μm) in deionized water were dispersed in a microchannel and sedimented on the planar glass substrate. The particles were actuated with a precessing magnetic field ($B_0$ = 10 mT, $\Omega$ = 18.8 rad/s, 3 Hz) (**Supplementary Note 1**). The precessing magnetic field is defined by two parameters: the precession angle $\Psi$, which is the angle between the instantaneous magnetic field vector $B(t)$ and the axis of precession, and the tilt angle $\vartheta$, which is the angle between the precession axis and the normal vector to the substrate plane **(Fig. 1a)**. Under the magnetic field, particles interact with their induced magnetic dipoles and form chains by head-to-toe alignment of their induced dipoles. Formation of chains is a dynamic self-assembly process, which depends on the magnetic field strength, the field frequency, the fluid viscosity, and the size and magnetic susceptibility of particles[8]. Once assembled, chains synchronously rotate with $B(t)$. By tilting the precession axis by $\vartheta$, chains self-propel over the substrate, orthogonal to the direction of the tilt. Self-propulsion results from the hydrodynamic symmetry breaking mechanism due to the rotation of chains near the solid boundary, which has been described in detail elsewhere[8,18,19].

We first investigated the pairwise dynamics of a homogeneous pair of chains for varying field parameters: $60° \leq \Psi \leq 76°$ at $\vartheta = 5°$ **(Figs. 1b, S1, Video S1)**. For $\Psi < 65°$, the trajectories of the pairs diverged, during which the distance between the pairs steadily grew until there was not any discernible pairwise interaction **(Figs. 1b, S1b)**. For $65° \leq \Psi \leq 72°$, the pairwise distance converged to and oscillated around a steady-state value that persisted during the experiment duration. In this regime, pairs translated and rotated around their center, which led to the trochoidal trajectories **(Fig. 1b, S1b)**. Increasing $\Psi$ further, the steady-state distance between the chains decreased until the pairs collapsed at $\Psi = 74°$ **(Fig. 1b, S1b)**. Increasing the number of chains, chains self-organized into a cluster with a discernible order in their spatial organization, where each chain was distanced at an approximately equal steady-state distance from their nearest neighbors **(Fig. 1c, Video S1)**. Clusters could be steered by changing the orientation of the tilt axis in the $x$-$y$ plane without changing $\Psi$ and $\vartheta$ **(Fig. 1c)**.

**Pairwise magnetic and hydrodynamic interactions**

Pairwise steady-state distances obtained from experiments are reported for different $\Psi$ and categorized under "divergence", "cohesion", and "collision" states based on their pairwise behavior (**Fig. 2a**). We attribute the emergence of a pairwise steady-state distance to dipolar and multipolar magnetic interactions between chains caused by their anisotropic shape magnetization. The magnetic interaction forces between chains were calculated numerically for all combinations of $\Psi$ and $\vartheta$, and for different number of microparticles per chain, $n$ (**Figs. 2a-b, S2, see Supplementary Note 2**). Particles that have isotropic magnetizations, such as spherical paramagnetic particles ($n = 1$), interact solely with their dipoles. For $\vartheta = 0°$, time-averaged dipolar interactions are repulsive when $\Psi$ is smaller than the magic angle 54.7°[20,21], and attractive for $\Psi > 54.7°$. In contrast, chains ($n > 1$) additionally have higher-order magnetic moments due to spatially separated dipoles positioned at centers of particles that form the chains[20,22,23]. For doublets ($n = 2$), the dominant repulsive multipolar interaction is the hexapole-dipole for $\Psi < 61.5°$, which decays rapidly with $r^{-6}$ compared to the $r^{-4}$ decay rate of the attractive dipolar interaction forces[20]. Similar multipolar interactions are also in play for chains with $n > 2$ (**Fig. S2e-g**). Due to the different decay rates of attractive dipolar and repulsive multipolar interactions, chains attract each other at large separations, and repel at small separations. The cross-over distance between the dipolar attraction and multipolar repulsion defines a dynamic steady-state distance $r^*$ between a pair of chains, and this distance depends on $\Psi$ and $n$ (*i.e.*, the length of the chain $L = 2an$) (**Figs. 2b, S2d-g**). For $\vartheta = 0°$, pairwise magnetic interaction is axisymmetric about the precession axis, with a negligible anisotropy produced by a small tilt of $\vartheta < 5°$ (**Fig. S2d**).

The pairwise motion of chains are mainly determined by the hydrodynamic interactions, especially when they are in the cohesive regime. Precessing chains generate fluid flows, which lead to their self-propulsion and to hydrodynamic interactions influencing the motion of their neighbors. Simulations were used for calculating the velocity field around precessing chains (**Supplementary**

**Note 3)**. The chain rotation perpendicular to the substrate leads to a rotational flow in the $x - y$ plane, leading to rotation of pairs about their center (**Figs. 2c, S3a**). For $\vartheta > 0°$, rotation has a component parallel to the substrate, leading to the self-propulsion of chains and the generation of a secondary flow field that increases the velocity of neighboring chains via advection (**Fig. 2d, S3b-c**).

Simulations combining magnetic and hydrodynamic interactions of chains quantitatively captured the dynamics of pair motion (**Fig. 2e-h, Video S2, Supplementary Note 4**). In the experiments, translational and rotational velocities of pairs increased while pair distance decreased with $\Psi$, which was supported by simulations (**Fig. 2f-h**). Furthermore, our model also captured pair dynamics for chains of different lengths. For increasing $n$, pair distance and translational velocity increased and rotational velocity decreased (**Fig. S4**).

**Dynamics and organization of homogeneous and heterogeneous clusters**

Increasing the number of chains $N$, we observed their self-organization into clusters (**Fig. 3a, b**). For homogeneous clusters, each chain was distanced evenly from their nearest neighbors (**Fig. 3c Video S3**). The pairwise distance decreased with $\Psi$ (**Fig. 3c**). The pairwise distance also decreased from $N = 2$ to $N = 3$, but remained relatively unchanged with further increases in $N$ (**Fig. 3c**). We ascribe this to the increased mutual dipolar attraction when multiple chains are present, which compacts the cluster. Cluster velocity increased with $\Psi$ and $N$ (**Fig. 3d, Video S3**). The latter effect can be attributed to an enhancement of collective hydrodynamic advection resulting from the superposition of flows generated by $N$ chains.

To understand the effects of intragroup variations on the collective dynamics, we investigated clusters formed by heterogeneous members (**Fig. 4**). Dynamic self-assembly allows tuning the size distribution in small clusters ($N = 7$), enabling systematic investigation of heterogeneity. Group stability was evaluated based on two order metrics: rotational order parameter, $R$, and connectivity (**Supplementary Note 6**). $R$ measures the degree of coherence of rotational motion of chains around the cluster center. $R \to 1$ for perfectly coherent rotational motion, and $R \to 0$ for no rotation[24]. Connectivity is based on total magnetic interaction potential; thus, quantifies the strength of cohesive interactions that hold the chains together.

Three clusters of varying degrees of heterogeneity were formed (**Fig. 4a**). Cluster heterogeneity was measured as the standard deviation of the number of particles in chains ($\sigma_n$), while the average $\bar{n}$ was fixed at 3. Homogeneous clusters actuated at $\Psi = 68°$ and $\vartheta = 5°$ remained as a single cluster, in which chains preserved their ordered spatial organization over time with $R \sim 1$ (**Fig. 4b, c**). However, when heterogeneities were introduced, clusters were more likely to break down, with decreasing $R$ and connectivity over time as the chain interactions weakened (**Fig. 4b, c, Video S4**). Decreasing $\vartheta$ to $3°$ reduced the cluster velocity and re-established the $R$ and connectivity of the heterogeneous clusters (**Fig. 4b, c, Video S4**). Overall, heterogeneities cause a variance in chain velocities (**Fig. S5**) and weakens cohesive forces, resulting in cluster dissolution. By decreasing the velocity, the balance of these competing effects is shifted towards cohesion.

To elucidate the group formation and dissolution dynamics, we investigated the chain trajectories after subtracting the mean cluster translation and rotation, revealing the internal motion of individual chains (**Fig. 4d, Video S4, Supplementary Note 6**). For stable clusters, chain displacements were constrained around their mean internal positions, where they perform small oscillations, as quantified by their mean-squared displacement (MSD) curves (**Fig. 4e**). We observed that the amplitude of these positional fluctuations increased with increasing heterogeneity and $\vartheta$ (**Fig. 4f**). Positional fluctuations were bounded to a finite amplitude when clusters exhibited structural ordering, and grew indefinitely when the clusters were breaking apart (*i.e.*, Het2, Ψ68° - 95°) (**Fig. 4f**).

**Formation of large clusters**

Scalability of cohesive interactions enabled formation of large clusters (up to $N = 53$), which were formed by introducing more particles during the cluster formation (**Fig. 5, Video S5**). Global motion and internal fluctuations of such clusters were measured experimentally (**Fig. 5a, b, Video S6**). We observed that the mean neighbor distance did not vary considerably with $N$ (**Fig. S6**). On the other hand, cluster velocity increased (**Fig. 5c, Video S6**), continuing the trend observed for smaller clusters (**Fig. 3d**). We ascribe this observation to scaling effects associated with the collective hydrodynamic interactions with increasing cluster size. To assess this argument, we developed a reduced-order model that captures magnetic and hydrodynamic interactions in large clusters (**Video S7, Supplementary Note 5**). Briefly, we model the attractive-repulsive magnetic interactions with an effective pair interaction force, and the hydrodynamic interactions with rotlet singularities located above a solid wall (**Fig. S7**). The flow generated by chain rotation parallel to the substrate has a positive contribution to propulsion velocity of neighboring chains, thus increases cluster velocity with increasing number of chains (**Figs. 2d and S7a**). The reduced-order model captures the qualitative features of the cluster motion, including trochoidal motion (**Fig. S7c**), and changes in the cluster velocity with increasing number of chains (**Figs. 5c, S7d, e**).

The internal motion of clusters reveals a tendency towards large positional fluctuations of chains with increasing $N$ (**Fig. 5b, Video S6**). In small clusters ($N \leq 11$), chains perform small fluctuations around their mean internal positions, which remain relatively fixed over long times, indicative of a solid-like order (**Fig. 5b, d**). As cluster size grew ($N > 19$), chains started moving inside the cluster, while remaining confined within the cluster radius $R_C$ (**Fig. 5b, d**). As such, amplitude of positional fluctuations grew proportional to $R_C$ when $N > 19$ **(Fig. 5d)**. MSD data showed that the chain motion exhibits ballistic $\sim t^2$ behavior at short times **(Fig. 5e)**. For $t > 10$ s, chain motion varied from small bounded oscillations evidenced in solid-like structuring in clusters for $N \leq 11$, to liquid-like diffusive $\sim t$ behavior confined to the cluster radius for $N \geq 15$ **(Fig. 5e)**.

We ascribe the tendency towards larger fluctuations with increasing $N$ to the different distance-dependent decay rates of magnetic and hydrodynamic interactions. Magnetic dipolar interactions decay with $r^{-4}$, whereas the far-field of rotlet flow parallel to the wall decays with $r^{-2}$ [25,26]. Being short-ranged, magnetic forces holding chains together in a solid-like order rapidly reach saturation as the number of neighboring chains increases. On the other hand, hydrodynamic forces displacing chains keep increasing due to their longer range. It can be expected that hydrodynamic effects compete magnetic effects at a certain cluster size, and the cluster transitions to a liquid-like internal state.

**Discussion**

Collective motion manifests itself at all length scales relevant to biology[27]. Examples range from cytoplasmic transport in plant cells[28], maze-solving slime moulds[29], cooperative foraging in social ant groups[30], migratory flocking of white storks[31], and human motion in crowded environments[32]. Inspired from natural systems, robotic swarm systems are being developed to address complex tasks such as collective construction and search operations[33,34]. A similar trend is prevalent in the microscale robotic swarms, with the aim of enhancing functional throughput, multitasking capabilities, and to impart microrobots with reconfigurability to enhance their adaptability to environmental constraints. However, a direct transfer of algorithmic approaches from the macroscale to microrobotic swarms present significant challenges due to challenges in miniaturization and powering of analogous components. Instead, microrobotic systems currently need to rely on micron-scale physical interactions for local coupling that generate global collective behaviors. Here, we demonstrated a microrobotic system where pairwise magnetic interactions can guide self-organization of cohesive clusters, which leads to a synergistic enhancement of the collective mobility over individual microrobots via hydrodynamic interactions.

Cohesive interactions play an important role in biological swarms (imagine a herd of sheep), where the group formation needs to be maintained without confining boundaries[15-17]. As opposed to purely attractive interactions that lead to the collapse of agents into tightly packed clusters, inclusion of short-range repulsive interactions can promote inter-agent spacing. In microrobotic swarms, several works have achieved cohesive organization in non-propulsive systems by combining repulsive capillary/hydrodynamic and attractive magnetic forces, however only at the liquid-air interfaces[35,36]. There is a great interest to form self-organizing swarms in fully liquid environments, due to their relevance to biomedical applications. In this work, we have achieved cohesive self-organization of microrobotic swarms in a fully immersed liquid environments by taking advantage of multipolar interactions resulting from anisotropic shape-magnetization of chains actuated with time-varying fields. Programming multipolar interactions holds promise for the design of advanced collective motion and manipulation of microrobotic swarm systems[6,8,22,23,37,38], as we have shown here.

Swarm heterogeneity can enhance resilience of social collectives against random noise[39], or can allow collaborative task division in robotic collectives[40]. On the other hand, introduction of large variances in population characteristics (*e.g.*, velocity, interaction strength) may have deleterious effects on the order and cohesiveness of flocking swarms that interact locally, which can be regulated or alleviated by self-sorting and mixing mechanisms[41-43]. Despite its importance, heterogeneity has remained relatively unexplored in the field of synthetic active matter swarms. In the present system, heterogeneities mainly contributed to disordering of cohesive clusters due to the variance of individual mobilities. However, collective order can be restored by slowing down the swarm, which re-adjusts the competing effects of cohesiveness and individuality.

Self-organized cohesion can be further scaled to form large swarms. In our system, we observed that increasing group size enhances swarm mobility, akin to hydrodynamic cooperation observed in collectives of sperm cells[44] and of active colloidal rollers[26,45,46]. On the other hand, different scaling of magnetic (decaying with $r^{-4}$, short range) and hydrodynamic (decaying with $r^{-2}$, long range) interactions affect the spatiotemporal organization of chains as swarms grow larger. We observed a solid-like spatial organization in small clusters, where magnetic cohesion is dominant over hydrodynamics. Two effects became comparable in larger groups, resulting in liquid-like dynamics, where chains displaced with respect to their neighbors. These states are highly similar to the flying crystal and the moving droplet formations in flocks interacting with cohesive alignment rules[16].

In conclusion, we have presented mobile microrobotic swarms cooperating to generate a cohesive organization much larger than an individual microrobotic unit, which introduce synergistic advantages and display rich spatiotemporal organizations arising from collective dynamics. Our approach addresses the operation of microrobots as localized teams, which could inspire researchers in active matter and microrobotics fields for designing advanced collective systems for future applications in biomedicine, precision manipulation and manufacturing, and environmental sensing and remediation.

**Materials and Methods**

*Experimental Setup*

External magnetic fields required for self-assembly and actuation of chain microrobots were generated using a custom five-coil magnetic setup integrated on an inverted optical microscope (Axio Observer A1, Carl Zeiss) **(Fig. S8)**. The magnetic coil system was arranged to generate up to 20 mT in in-plane directions and 10 mT in out of plane z-direction[7]. All experiments were performed in a closed microfluidic channel (75 µm height x 6 mm width x 10 mm length) composed of laser micromachined poly(methyl methacrylate) (PMMA) tops with fluidic connections and double sided tape, defining the channel outline and height, assembled with a cover glass[47]. Superparamagnetic polystyrene microparticles (5 µm in diameter, Sigma Aldrich) were used in self-assembly of chain microrobots. The microparticles were suspended in a 0.1% Tween 20 (Sigma Aldrich, St Louis, MO) solution in deionized water and injected into the microchannel. A precessing magnetic field with pre-determined tilt and precession angles is applied to assemble and actuate chain microrobots, which is detailed in **Supplementary Note 1**.

*Dynamic Model and Simulations*

Magnetic interactions between two chains were calculated by averaging dipolar interaction forces between particles over a cycle of magnetic field precession. We defined a characteristic magnetic interaction force, $F_0 = \frac{\pi}{12\mu_0}\left(\frac{a\chi B}{n}\right)^2$, for two chains separated by a distance *L*. Detailed methods for calculating magnetic interactions are described in **Supplementary Note 2**.

Hydrodynamic simulations were used for calculating the fluid flow induced by the motion of a precessing chain. For given particle kinematics, flow velocity at an arbitrary point in space can be calculated via the hydrodynamic Blake tensor, which accounts for the no-slip boundary condition at the wall surface[25]. Method for calculating the flow fields is detailed in **Supplementary Note 3**.

Pairwise dynamics of chains was simulated by modeling the motion of individual particles constituting the chains. The model takes into account magnetic, hydrodynamic, and excluded volume interactions between particles that are driven by a precessing magnetic field. The equation of motion for particles is given by:

$$\dot{R}_i = \mathcal{M}_{ij} \cdot (f_{m,j} + f_{b,j} + f_{w,j} + f_{g,j}) \tag{Eq. 1}$$

where interaction forces between induced magnetic dipoles of particles ($f_m$), particle-particle ($f_b$) and particle-wall ($f_w$) excluded volume forces, and gravitational ($f_g$) forces are calculated with appropriate physical models. The grand mobility tensor ($\mathcal{M}$) is used for calculating velocities ($\dot{R}_i$) of hydrodynamically interacting particles near a no-slip boundary, for a set of forces acting on the ensemble of particles[48]. Details of the dynamic model are described in **Supplementary Note 4**.

A reduced-order discrete chain model was developed for simulating the dynamics of ensembles of chains that self-organize into mobile clusters. Briefly, the model considers each chain as a discrete agent, and uses time-averaged force functions for calculating their magnetic and hydrodynamic interactions. Details of the model are described in **Supplementary Note 5**.

*Data Analysis*

Acquired microscopy images were processed using Fiji[49] to identify individual chains and their position. A tracking software[50] was used to generate trajectories of chains and to determine their velocities. All data analysis was performed on MATLAB (MathWorks, Inc.). Methods to analyze cluster translation, rotation and internal motion of chains are described in **Supplementary Note 6**.

**Acknowledgements**

Y.A. thanks Alexander von Humboldt Foundation for the Humboldt Postdoctoral Research Fellowship. This work is funded by the Max Planck Society.

**Competing financial interests**

The authors declare no competing financial interests.

**References**

1. Sitti M. *Mobile Microrobotics*. MIT Press, Cambridge, MA (2017).
2. Alapan Y, Yasa O, Yigit B, Yasa IC, Erkoc P, Sitti M. Microrobotics and Microorganisms: Biohybrid Autonomous Cellular Robots. *Annual Review of Control, Robotics, and Autonomous Systems*.
3. Erkoc P, Yasa IC, Ceylan H, Yasa O, Alapan Y, Sitti M. Mobile microrobots for active therapeutic delivery. *Advanced Therapeutics* **2**, 1800064 (2019).
4. Zhang Z, Wang X, Liu J, Dai C, Sun Y. Robotic Micromanipulation: Fundamentals and Applications. *Annual Review of Control, Robotics, and Autonomous Systems* **2**, 181-203 (2019).
5. Soler L, Sánchez S. Catalytic nanomotors for environmental monitoring and water remediation. *Nanoscale* **6**, 7175-7182 (2014).
6. Alapan Y, Yigit B, Beker O, Demirors AF, Sitti M. Shape-encoded dynamic assembly of mobile micromachines. *Nature Materials*, doi: 10.1038/s41563-019-0407-3 (2019).
7. Alapan Y, *et al.* Soft erythrocyte-based bacterial microswimmers for cargo delivery. *Science Robotics* **3**, (2018).
8. Yigit B, Alapan Y, Sitti M. Programmable Collective Behavior in Dynamically Self-Assembled Mobile Microrobotic Swarms. *Advanced Science* **6**, 1801837 (2019).
9. Martel S, Mohammadi M. Using a swarm of self-propelled natural microrobots in the form of flagellated bacteria to perform complex micro-assembly tasks. *2010 IEEE International Conference on Robotics and Automation*, 500-505, IEEE, (2010).
10. Servant A, Qiu F, Mazza M, Kostarelos K, Nelson BJ. Controlled In Vivo Swimming of a Swarm of Bacteria-Like Microrobotic Flagella. *Advanced Materials* **27**, 2981-2988 (2015).
11. Yu J, Wang B, Du X, Wang Q, Zhang L. Ultra-extensible ribbon-like magnetic microswarm. *Nature Communications* **9**, 3260 (2018).
12. Wang B, *et al.* Reconfigurable Swarms of Ferromagnetic Colloids for Enhanced Local Hyperthermia. *Advanced Functional Materials* **28**, 1705701 (2018).
13. Xie H, *et al.* Reconfigurable magnetic microrobot swarm: Multimode transformation, locomotion, and manipulation. *Science Robotics* **4**, eaav8006 (2019).
14. Parrish JK, Edelstein-Keshet L. Complexity, Pattern, and Evolutionary Trade-Offs in Animal Aggregation. *Science* **284**, 99-101 (1999).
15. Couzin ID, Krause J, James R, Ruxton GD, Franks NR. Collective memory and spatial sorting in animal groups. *Journal of Theoretical Biology* **218**, 1-11 (2002).
16. Grégoire G, Chaté H, Tu Y. Moving and staying together without a leader. *Physica D: Nonlinear Phenomena* **181**, 157-170 (2003).
17. Lavergne FA, Wendehenne H, Bauerle T, Bechinger C. Group formation and cohesion of active particles with visual perception-dependent motility. *Science* **364**, 70-74 (2019).

18. Sing CE, Schmid L, Schneider MF, Franke T, Alexander-Katz A. Controlled surface-induced flows from the motion of self-assembled colloidal walkers. *Proceedings of the National Academy of Sciences* **107**, 535-540 (2010).

19. Zhang L*, et al.* Controlled Propulsion and Cargo Transport of Rotating Nickel Nanowires near a Patterned Solid Surface. *ACS Nano* **4**, 6228-6234 (2010).

20. Tierno P, Schreiber S, Zimmermann W, Fischer TM. Shape Discrimination with Hexapole−Dipole Interactions in Magic Angle Spinning Colloidal Magnetic Resonance. *Journal of the American Chemical Society* **131**, 5366-5367 (2009).

21. Salehizadeh M, Diller E. Two-agent formation control of magnetic microrobots in two dimensions. *Journal of Micro-Bio Robotics* **12**, 9-19 (2017).

22. Tierno P. Recent advances in anisotropic magnetic colloids: realization, assembly and applications. *Physical Chemistry Chemical Physics* **16**, 23515-23528 (2014).

23. Klapp SH. Collective dynamics of dipolar and multipolar colloids: From passive to active systems. *Current Opinion in Colloid & Interface Science* **21**, 76-85 (2016).

24. Attanasi A*, et al.* Collective behaviour without collective order in wild swarms of midges. *PLoS Computational Biology* **10**, e1003697 (2014).

25. Blake JR, Chwang AT. Fundamental singularities of viscous flow. *Journal of Engineering Mathematics* **8**, 23-29 (1974).

26. Driscoll M, Delmotte B, Youssef M, Sacanna S, Donev A, Chaikin P. Unstable fronts and motile structures formed by microrollers. *Nature Physics* **13**, 375 (2017).

27. Vicsek T, Zafeiris A. Collective motion. *Physics Reports* **517**, 71-140 (2012).

28. Woodhouse FG, Goldstein RE. Cytoplasmic streaming in plant cells emerges naturally by microfilament self-organization. *Proceedings of the National Academy of Sciences* **110**, 14132-14137 (2013).

29. Nakagaki T, Yamada H, Tóth Á. Intelligence: Maze-solving by an amoeboid organism. *Nature* **407**, 470 (2000).

30. Feinerman O, Pinkoviezky I, Gelblum A, Fonio E, Gov NS. The physics of cooperative transport in groups of ants. *Nature Physics*, 1 (2018).

31. Flack A, Nagy M, Fiedler W, Couzin ID, Wikelski M. From local collective behavior to global migratory patterns in white storks. *Science* **360**, 911-914 (2018).

32. Silverberg JL, Bierbaum M, Sethna JP, Cohen I. Collective motion of humans in mosh and circle pits at heavy metal concerts. *Physical Review Letters* **110**, 228701 (2013).

33. Werfel J, Petersen K, Nagpal R. Designing collective behavior in a termite-inspired robot construction team. *Science* **343**, 754-758 (2014).

34. Stirling T, Wischmann S, Floreano D. Energy-efficient indoor search by swarms of simulated flying robots without global information. *Swarm Intelligence* **4**, 117-143 (2010).

35. Wang W, Giltinan J, Zakharchenko S, Sitti M. Dynamic and programmable self-assembly of micro-rafts at the air-water interface. *Science Advances* **3**, e1602522 (2017).

36. Wang Q, Yang L, Wang B, Yu E, Yu J, Zhang L. Collective Behavior of Reconfigurable Magnetic Droplets via Dynamic Self-Assembly. *ACS Applied Materials & Interfaces* **11**, 1630-1637 (2019).

37. Miyashita S, Diller E, Sitti M. Two-dimensional magnetic micro-module reconfigurations based on inter-modular interactions. *The International Journal of Robotics Research* **32**, 591-613 (2013).

38. Diller E, Pawashe C, Floyd S, Sitti M. Assembly and disassembly of magnetic mobile micro-robots towards deterministic 2-D reconfigurable micro-systems. *The International Journal of Robotics Research* **30**, 1667-1680 (2011).

39. Miguel MC, Parley JT, Pastor-Satorras R. Effects of heterogeneous social interactions on flocking dynamics. *Physical Review Letters* **120**, 068303 (2018).

40. Ducatelle F, Di Caro GA, Pinciroli C, Gambardella LM. Self-organized cooperation between robotic swarms. *Swarm Intelligence* **5**, 73 (2011).

41. Copenhagen K, Quint DA, Gopinathan A. Self-organized sorting limits behavioral variability in swarms. *Scientific Seports* **6**, 31808 (2016).

42. Yllanes D, Leoni M, Marchetti M. How many dissenters does it take to disorder a flock? *New Journal of Physics* **19**, 103026 (2017).

43. Fu X*, et al.* Spatial self-organization resolves conflicts between individuality and collective migration. *Nature communications* **9**, 2177 (2018).

44. Yang Y, Elgeti J, Gompper G. Cooperation of sperm in two dimensions: Synchronization, attraction, and aggregation through hydrodynamic interactions. *Physical Review E* **78**, 061903 (2008).

45. Martinez-Pedrero F, Ortiz-Ambriz A, Pagonabarraga I, Tierno P. Colloidal Microworms Propelling via a Cooperative Hydrodynamic Conveyor Belt. *Physical Review Letters* **115**, 138301 (2015).

46. Driscoll M, Delmotte B. Leveraging collective effects in externally driven colloidal suspensions: Experiments and simulations. *Current opinion in colloid & interface science*, (2018).

47. Alapan Y, Matsuyama Y, Little JA, Gurkan UA. Dynamic deformability of sickle red blood cells in microphysiological flow. *Technology* **4**, 71-79 (2016).

48. Swan JW, Brady JF. Simulation of hydrodynamically interacting particles near a no-slip boundary. *Physics of Fluids* **19**, 113306 (2007).

49. Schindelin J*, et al.* Fiji: an open-source platform for biological-image analysis. *Nature Methods* **9**, 676 (2012).

50. Crocker JC, Grier DG. Methods of Digital Video Microscopy for Colloidal Studies. *Journal of Colloid and Interface Science* **179**, 298-310 (1996).

**Figures**

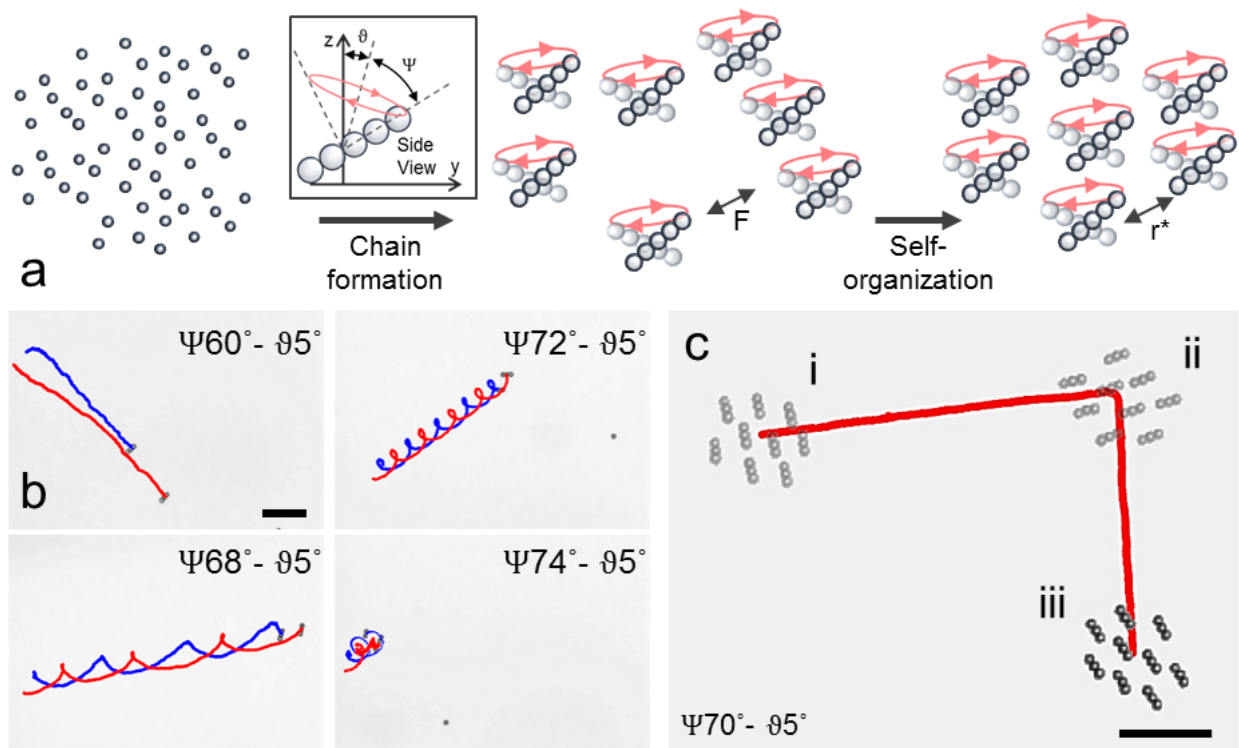

**Figure 1. Cohesive self-organization of microrobotic clusters. (a)** A collection of dispersed paramagnetic microparticles are actuated with a precessing magnetic field defined by angles Ψ and ϑ. Particles self-assemble into linear chains simultaneously in large numbers and interact with multipolar magnetic forces. Chains self-organize into cohesive clusters by arranging themselves at steady-state distances from their neighbors ($r^*$). **(b)** Experimental trajectories for a pair of chains with $n = 3$ (*n*: number of particles per chain). Pairwise distance between chains diverge at Ψ60º − ϑ5º, converge to a steady-state distance at Ψ68º − ϑ5º and Ψ72º − ϑ5º, and chains collapse at Ψ74º − ϑ5º. **(c)** Group formation and steering of a cluster formed by nine chains. Scale bars are 50 µm.

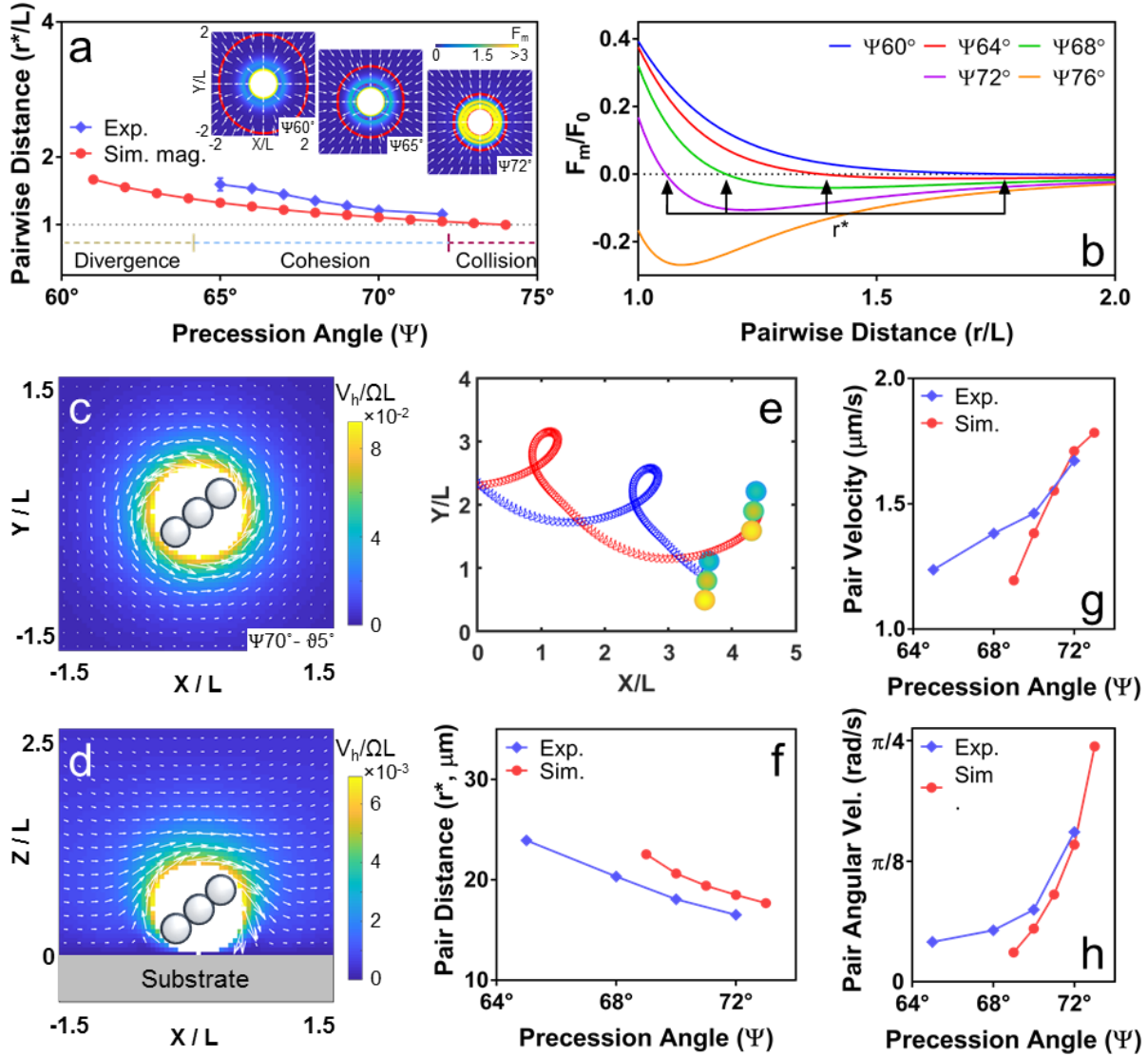

**Figure 2. Pairwise magnetic and hydrodynamic interactions. (a)** Experimentally measured pairwise steady-state distances ($r^*$) are compared to simulations of magnetically interacting chains at different $\Psi$. Insets show simulations of magnetic interaction forces between two chains varying with their relative positions. Arrows indicate the direction of magnetic force. Red curve indicates the steady-state distance arising from magnetic interactions. Colorbar indicates the magnitude of magnetic force, $F_m$. Forces are normalized to $F_0$. Error bars indicate the standard deviations obtained over three separate experiments. **(b)** Magnetic interaction force varying with pairwise distance ($r$) at different $\Psi$. $F_m < 0$ attracts and $F_m > 0$ repels. The cross-over distance gives the steady-state distance, $F_m(r^*) = 0$. **(c, d)** Simulations of hydrodynamic fields induced by a precessing chain, visualized in planes **(c)** parallel and **(d)** perpendicular to the substrate. Colorbars indicate flow velocity $V_h$. **(e-h)** Simulations of magnetically and hydrodynamically interacting chains reproduce **(e)** experimental trajectories, **(f)** steady-state distance, **(g)** pair translation and **(h)** angular velocity at different $\Psi$. Simulations were performed for chains with 3 particles and for $\vartheta = 5°$. $L$ denotes the chain length.

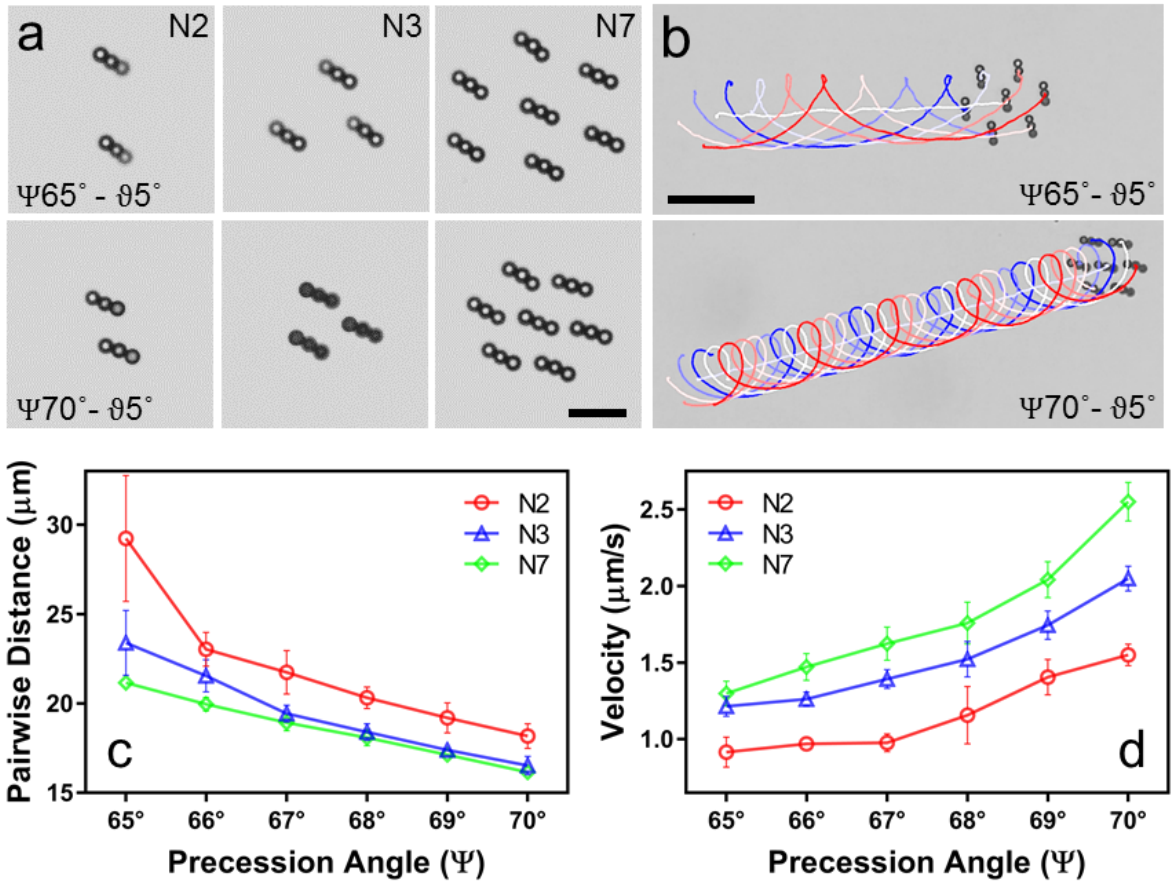

**Figure 3. Self-organization and motion of homogeneous clusters.** (**a**) Experimental snapshots of clusters that were formed by $N$ = 2, 3, and 7 chains under two different precession angles $\Psi$ = 65° and 70° at a fixed tilt angle, $\vartheta$ = 5°. Scale bar is 20 μm. (**b**) Clusters translated and rotated along a straight line that led to the trochoidal trajectories of chains. Scale bar is 50 μm. (**c**) Pairwise distance between neighboring chains decreased with increasing $N$ and $\Psi$. (**d**) Cluster velocity increased with $N$ and $\Psi$. Error bars indicate the standard deviations obtained over three separate experiments.

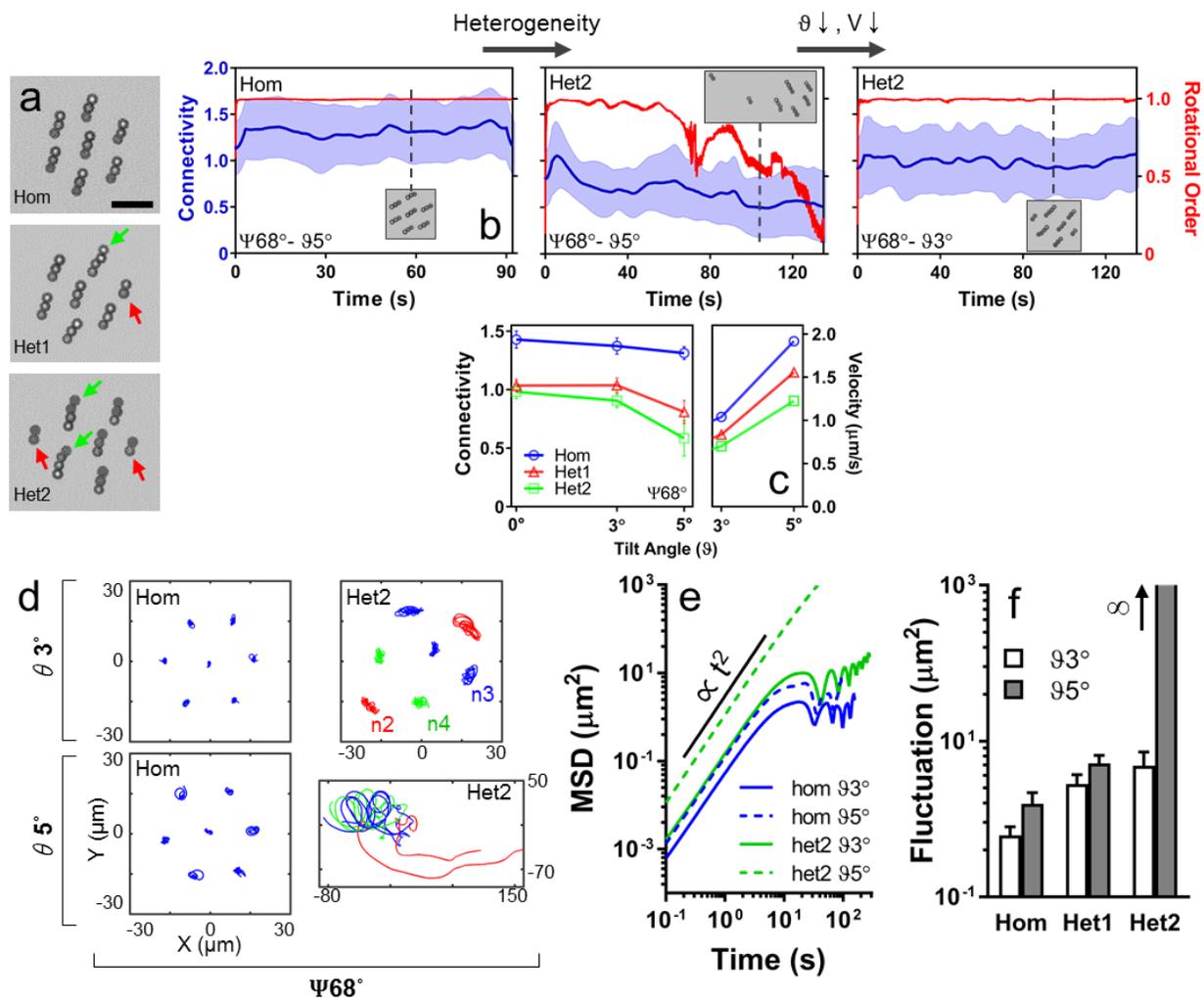

**Figure 4. Motion and internal dynamics of heterogeneous clusters. (a)** Experimental snapshots of three clusters with different levels of heterogeneities. Heterogeneity is measured as the standard deviation of number of beads in chains ($\sigma_n$) where the mean $\bar{n}$ is fixed at 3. Cluster heterogeneity increases from Hom ($\sigma_n = 0$) to Het1 ($\sigma_n = 0.58$) and Het2 ($\sigma_n = 0.82$). Hom: homogeneous Het: heterogeneous. Scale bar is 25 μm. **(b)** Increasing heterogeneity, rotational order and connectivity of the cluster decrease in time as the cluster dissolves. Decreasing the tilt angle $\vartheta$ reduces the cluster velocity $V$ and re-establishes the rotational order and cohesiveness of the heterogeneous cluster. Insets show experimental snapshots at indicated time points. **(c)** Average connectivity decreased

with increasing heterogeneity and $\vartheta$, whereas the cluster velocity decreased with decreasing $\vartheta$ and increasing heterogeneity. Error bars indicate the standard deviations obtained over time-series measurements. **(d)** Internal positional fluctuations of the chains are revealed after subtracting the cluster translation and rotation. Colors of the trajectories indicate the number of particles (red, $n = 2$, blue, $n = 3$, green, $n = 4$). **(e)** Mean-squared displacement (MSD) data show that the chains were constrained around their mean internal position for stable clusters, indicated by the long time plateau at $t > 10$ s for all cases except for Het2 $\vartheta 5°$. **(f)** Positional fluctuation amplitude of chains increased with heterogeneity and $\vartheta$. Error bars indicate the standard deviations averaged over all chains.

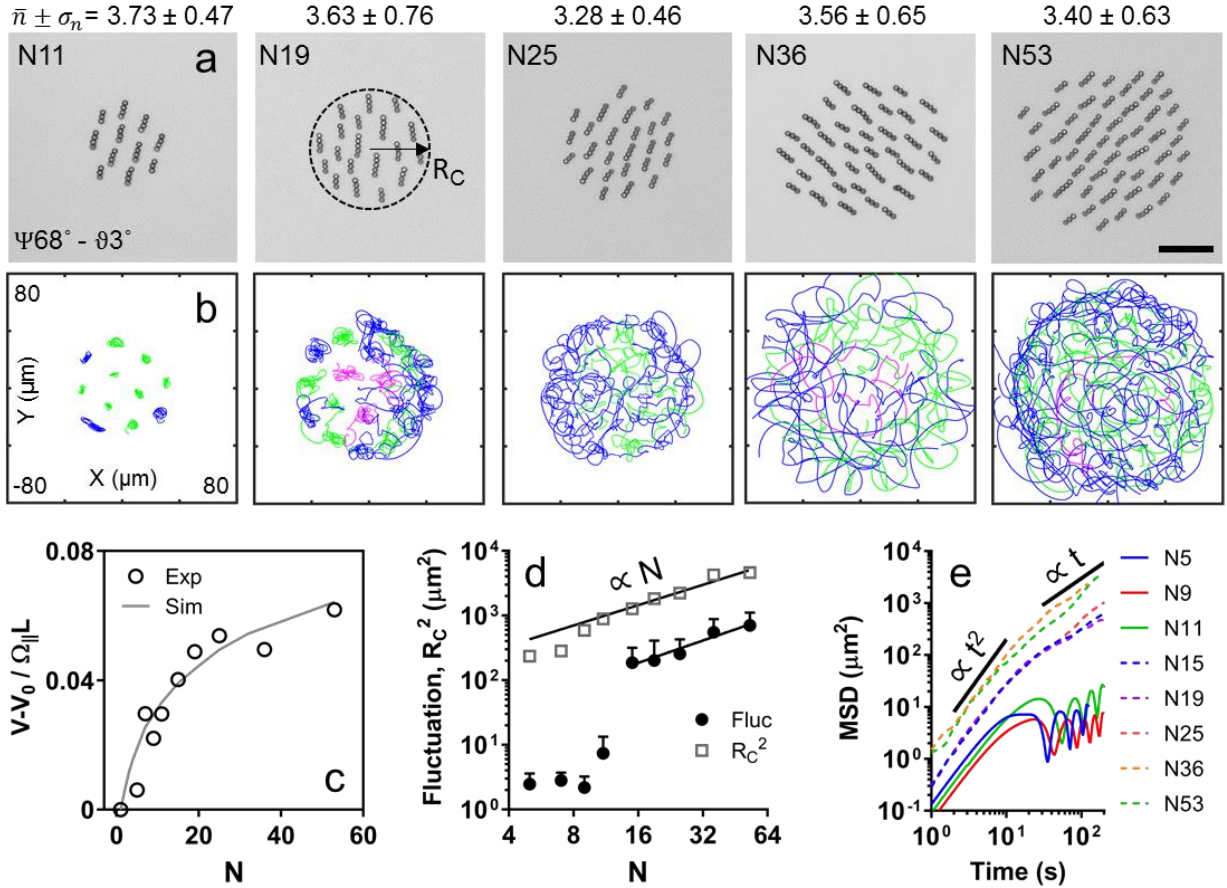

**Figure 5. Motion and internal dynamics of large clusters. (a)** Experimental snapshots of clusters with increasing number of chains (*N*) actuated at Ψ68° - ϑ3°. Mean ± standard deviation of number of beads in chains (*n*) are shown on top of the images. $R_C$ is the measured radius of the cluster. Scale bar is 50 μm. **(b)** Trajectories of chains after subtracting cluster translation and rotation reveal their positional fluctuations. Colors of the trajectories indicate the number of particles in the corresponding chains (blue, *n* = 3, green, *n* = 4, magenta, *n* = 5). In small clusters (*N* < 15), chains perform small displacements around their mean internal positions, which is indicative of a solid-like order. As cluster size grows individual chains start displacing inside the cluster, while remaining confined within the cluster radius, which is indicative of a liquid-like behavior. **(c)** Cluster velocity relative to the velocity of a single chain (V-V$_0$) increases with cluster size, which is predicted by simulations. **(d)** $R_C$ increases approximately with $\sqrt{N}$, indicative of constant density

of chains (*i.e.*, constant neighbor distances). Amplitude of positional fluctuations grow proportional to $R_C$ with increasing $N$ for $N \geq 15$. Error bars indicate standard deviations averaged over all chains. **(e)** MSD curves show ballistic $\sim t^2$ behavior for the chain motion at short times $t < 10$ s. At long times, MSD curves show that the chains exhibit solid-like behavior for N < 15 and liquid-like $\sim t$ behavior for N $\geq$ 15.

# Cohesive self-organization of mobile microrobotic swarms

Berk Yigit[#], Yunus Alapan[#], and Metin Sitti[*]

Physical Intelligence Department, Max Planck Institute for Intelligent Systems, 70569 Stuttgart, Germany

[*] E-mail: sitti@is.mpg.de

[#] The authors contributed equally and share the first authorship.

**Supplementary Note 1: Applied magnetic fields**

Precessing magnetic field is defined by two angles. Precession angle $\Psi$ is the angle between the axis of precession ($w$) and the magnetic field vector ($B$), and the tilt angle $\vartheta$ is the angle between the precession axis and the normal vector to the planar substrate surface (*i.e.*, z-axis) (Fig. 1a). Applied magnetic field is varied in time by revolving the magnetic field vector about the precession axis, with angular velocity $\Omega$ via the following mathematical operation:

$$\boldsymbol{B}(t) = B_0(\boldsymbol{I} + \sin(\Omega t)\,[\boldsymbol{w}]_\times + (1-\cos(\Omega t))[\boldsymbol{w}]_\times^2)\boldsymbol{n}_0 \qquad \text{(Eq. S1)}$$

where $B_0 = \|\boldsymbol{B}\|$ is the magnetic field magnitude, $t$ is time, $\boldsymbol{I}$ is the identity matrix, $[\ ]_\times$ is the cross-product operator, $\boldsymbol{w} = (0, \sin(\vartheta), \cos(\vartheta))$, and $\boldsymbol{n}_0 = (0, \sin(\vartheta + \Psi), \cos(\vartheta + \Psi))$ is the magnetic field vector at $t = 0$.

**Supplementary Note 2: Magnetic interactions between chains**

For calculation of magnetic interactions forces between two chains (Figs. 2a-b and S2), we considered that each chain consists of $n$ paramagnetic particles and precess by following the magnetic field given by Eq. S1. Distance vector pointing from the $j^{th}$ particle to $i^{th}$ particle is denoted by $\boldsymbol{R}_{ij} = \boldsymbol{R}_j - \boldsymbol{R}_i$. The induced magnetic dipole moment $\boldsymbol{m}_i$ of each particle is given by $\boldsymbol{m}_i = v_p \chi \boldsymbol{B}/\mu_0$, where $v_p$ is the particle volume, $\chi$ is the volumetric magnetic susceptibility and $\mu_0$ is the vacuum permeability. The interaction force between two magnetic dipoles is calculated with the following equation[1],

$$\boldsymbol{f}_{m,ij} = \frac{3\mu_0}{4\pi R^4}\left(\frac{\boldsymbol{R}_{ij}(\boldsymbol{m}_i.\boldsymbol{m}_j)+\boldsymbol{m}_i(\boldsymbol{R}_{ij}.\boldsymbol{m}_j)+\boldsymbol{m}_j(\boldsymbol{R}_{ij}.\boldsymbol{m}_i)}{R} - \frac{5\boldsymbol{R}_{ij}(\boldsymbol{m}_i.\boldsymbol{R}_{ij})(\boldsymbol{m}_j.\boldsymbol{R}_{ij})}{R^3}\right) \quad \text{(Eq. S2)}$$

Following, the time-averaged magnetic force acting between two chains can be calculated by summing the interaction forces between particles and averaging over a precession cycle,

$$\boldsymbol{F}_m = \omega \int_0^{1/\omega} \sum_{i \neq j} \boldsymbol{f}_{m,ij}(t)\, dt \quad \text{(Eq. S3)}$$

where the summation is performed over each particle pair $i, j$ belonging to chains. Observing Eqs. S2 and S3, the strength of the magnetic force between two chains varies with the following proportionality:

$$F_m \propto n^2 \left(\frac{4\pi}{3\mu_0}\frac{(a^3\chi B)^2}{r^4}\right) \quad \text{(Eq. S4)}$$

where $r$ is the distance between two chains. Therefore, for a pair of chains separated by a distance of one chain length ($r = 2na$), a characteristic magnetic interaction force can be defined as:

$$F_0 = \left(\frac{\pi}{12\mu_0}\frac{(a\chi B)^2}{n^2}\right) \quad \text{(Eq. S5)}$$

**Supplementary Note 3: Numerical model for simulating flow fields around a single chain**

Flow field generated by the precession of a chain near a wall was calculated with simulations using Stokes flow singularities. Flow velocity at a given position in space $\boldsymbol{v}_h(\boldsymbol{r})$ due to a collection of particles on which a force $\boldsymbol{f}_j(\boldsymbol{R}_j)$ (*i.e.*, a Stokeslet) acts can be obtained with the following equation,

$$v_h(r) = \sum_j \mathbb{J}(r, R_j) \cdot f_j(R_j) \quad \text{(Eq. S6)}$$

In an unbounded fluid, $\mathbb{J}(r, R_j)$ is given by the Oseen tensor,

$$\mathbb{J}(r, R_j) = \frac{1}{8\pi\mu}\left(\frac{I}{|r-R_j|} + \frac{(r-R_j)(r-R_j)}{|r-R_j|^3}\right) \quad \text{(Eq. S7)}$$

Where $\mu$ is the dynamic viscosity of the fluid. Wall effects can be accounted for by modifying Eq. S7 through the image system of hydrodynamic singularities satisfying no-slip boundary conditions at the surface, which was formulated in the work of Blake and Chwang[2] and was also used in our simulations. Flow velocities presented in Figs. 2c-d and Fig. S3 were obtained by calculating time-varying flow field generated by a chain and taking its average over a precession cycle.

**Supplementary Note 4: Numerical model for simulating the chain dynamics**

For simulating the dynamics of motile self-assembled chains (Figs. 2e-h and S4, Video S2), we model the dynamics of the collection of particles that constitute the chains, whose equation of motion is given by,

$$\dot{R}_i = \mathcal{M}_{ij} \cdot (f_{m,j} + f_{b,j} + f_{w,j} + f_{g,j}) \quad \text{(Eq. S8)}$$

where interactions between particles $i$ and $j$ arise from magnetic dipole-dipole forces ($f_m$), particle-particle ($f_b$) and particle-wall ($f_w$) excluded volume forces, and gravitational ($f_g$) forces. Magnetic dipolar interactions between particles are calculated via Eq. S2. The grand mobility tensor $\mathcal{M}$ couples the velocities of particles ($\dot{R}_i$) to the forces acting on each particle through contributions of self and pair hydrodynamic mobility tensors that account for no-slip boundary conditions at the substrate surface[3]. Simulations implement the mathematical expressions for the grand mobility tensor that were provided by Swan and Brady[3]. Following the approach presented by Sing et al.[4], particle-particle and particle-wall excluded volume forces were modeled with modified Lennard-Jones force terms,

$$f_b = \frac{\epsilon}{r-2a}\left(\left(\frac{\sigma}{r-2a}\right)^{12} - \left(\frac{\sigma}{r-2a}\right)^{6}\right) \qquad \text{(Eq. S9)}$$

$$f_w = \frac{\epsilon}{h-a}\left(\left(\frac{\sigma}{h-a}\right)^{12} - \left(\frac{\sigma}{h-a}\right)^{6}\right) \qquad \text{(Eq. S10)}$$

with σ = 0.1$a$, $h$ is the distance of a particle from the wall, and $\epsilon$ is sufficiently small to neglect attractive terms. $f_g$ = - $\Delta\rho v_p g$ where $\Delta\rho$ is the buoyant density of particles, and $g$ is the gravitational acceleration. Eq. S8 was integrated with an explicit Euler scheme to obtain the trajectories of each particle in chains. Each cycle of chain rotation was divided into $5\times10^5$ time steps, and simulations were performed for 270 cycles (~30 seconds of real time experiments).

**Table S1.** Simulation parameters

| $a$ | 2.5 μm | $\chi$ | 0.5 |
|---|---|---|---|
| $B_0$ | 10 mT | $\Delta\rho$ | 0.05 g/cm$^3$ |
| $\Omega/2\pi$ | 3 Hz | $\mu$ | 0.894 mPa.s |

**Supplementary Note 5: Reduced-order discrete chain model**

For simulating the dynamics of clusters consisting of many chains, calculating the motion of each particle separately is computationally intensive. For this reason, we developed a reduced order simulation that models the dynamic of the collection of chains that constitute the cluster, in which each chain is treated as a discrete point (Fig. S6 and Video S7). This model accounts for the time-averaged magnetic interactions and near-wall hydrodynamic self-propulsion and interactions between chains. The equation of motion for each chain is given by

$$\dot{\boldsymbol{r}}_i = \boldsymbol{v}_{0,i} + \sum_{j \neq i} \boldsymbol{v}_{h,ij} + \sum_{j \neq i} \boldsymbol{v}_{m,ij} \qquad \text{(Eq. S11)}$$

where the velocity of $i^{th}$ chain ($\dot{\boldsymbol{r}}_i$), is the sum of its self-propulsion velocity, $\boldsymbol{v}_{0,i}$, the velocity of the hydrodynamic flow generated by all its $j^{th}$ neighbours at the position of $i^{th}$ chain, $\boldsymbol{v}_{h,ij}$, and the displacement velocity due to the magnetic forces imposed by its neighbors, $\boldsymbol{v}_{m,ij}$.

Self-propulsion of a chain arises from its self-advection under the hydrodynamic flow generated by its precessing motion. Similarly, a chain is also advected by the flow generated by its neighbors. The essential features of hydrodynamic flows were captured by modeling each chain as a rotlet singularity above a solid wall. Specifically, a rotlet solution provides the hydrodynamic velocity field generated by a point torque at Stokes regime. We consider that each chain has an effective hydrodynamic radius $a_h$, and its rotation is given with the angular velocity vector $\boldsymbol{\Omega}$. The axis of rotation is tilted from the $z$-axis (*i.e.*, normal to the plane of the substrate) by angle $\vartheta$, therefore, the angular velocity vector can be decomposed into two components which are parallel and perpendicular to the substrate, such that $\boldsymbol{\Omega} = (0, \Omega_\parallel, \Omega_\perp)$ where $\Omega_\parallel = \Omega \sin(\vartheta)$ and $\Omega_\perp = \Omega \cos(\vartheta)$. Chains are located at a distance $h$ from the wall at $\boldsymbol{r} = (x, y, h)$. The solid wall imposes hydrodynamic no-slip boundary conditions, which can be accounted through the image system consisting of a counter-rotating rotlet, and an additional stresslet and a source doublet positioned at $\boldsymbol{r}_{im} = (x, y, -h)$, which was formulated in the work of Blake and Chwang[2]. Thus, the flow velocity at the position of the $i^{th}$ chain is calculated by the following expressions[2,5]:

$$v_{0x,i} = \frac{\Omega_{\parallel,i} a_{h,i}^5}{8 h_i^4} \qquad \text{(Eq. S12)}$$

$$v_{hx,ij} = \Omega_{\|,j} a_{h,j}^3 \left( 6h_i \frac{(x_j-x_i)^2}{r_{im,ij}^5} - \frac{h_j-h_i}{r_{ij}^3} + \frac{h_j-h_i}{r_{im,ij}^3} \right) + \Omega_{\perp,j} a_{h,j}^3 \left( \frac{y_j-y_i}{r_{ij}^3} - \frac{y_j-y_i}{r_{im,ij}^3} \right) \quad \text{(Eq. S13)}$$

$$v_{hy,ij} = \Omega_{\|,j} a_{h,j}^3 \left( 6h_i \frac{(x_j-x_i)(y_j-y_i)}{r_{im,ij}^5} \right) + \Omega_{\perp,j} a_{h,j}^3 \left( \frac{x_i-x_j}{r_{ij}^3} - \frac{x_i-x_j}{r_{im,ij}^3} \right) \quad \text{(Eq. S14)}$$

where $r_{ij} = \left( (x_i-x_j)^2 + (y_i-y_j)^2 + (h_i-h_j)^2 \right)^{1/2}$ and $r_{im,ij} = \left( (x_i-x_j)^2 + (y_i-y_j)^2 + (h_i+h_j)^2 \right)^{1/2}$.

The first role of $\Omega_{\|}$ is self-propulsion, a chain rotating about the y-axis would translate in the x-direction, which is expressed in Eq. S12. The hydrodynamic velocity field generated by $\Omega_{\|}$ in the x - y plane is displayed in Fig. S6a. Flow field has a positive velocity in the x-direction around the rotlet. This leads to an enhancement of translation velocity in the x-direction for neighboring chains as a result of being advected by the flows generated by each other. The $\Omega_{\perp}$ component of rotation results in a rotating flow in the x - y plane (Fig. S6b), and is the main contributor to the rotation of chains around each other, and leads to the rotation of the cluster.

For calculating $\mathbf{v}_m$, we modeled the magnetic interaction force between chains with a time-averaged force that acts along the line connecting chain centers. Magnetic interaction force combines an attractive dipolar term and a repulsive multipolar term, which is given via the following equation:

$$\mathbf{F}_{m,ij} = \left[ \frac{A}{r^4} + \frac{B}{r^k} \right] \hat{\mathbf{r}} \quad \text{(Eq. S15)}$$

where $A$ and $B$ are the coefficients for dipolar and multipolar contributions, respectively, and $\hat{\mathbf{r}}$ is the unit vector pointing from chain $i$ to $j$. Dipolar interaction has a $1/r^4$ rate of decay, and multipolar interaction has an effective decay rate of $1/r^k$.

For determination of $A$, we consider the time-averaged effective dipolar interactions between two chains that precess about the $z$-axis with angle $\Psi$. The time averaged dipolar coupling strength under precession is given by[6]

$$A = \frac{3\pi}{4\mu_0} m_i m_j \left(\frac{3\cos\Psi - 1}{2}\right) \qquad \text{(Eq. S16)}$$

where the total dipole moment $\boldsymbol{m}_i$ is given by $\boldsymbol{m}_i = n\boldsymbol{m}$, where $n$ is the number of particles in the chain and $\boldsymbol{m}$ is the magnetic dipole moment of a single particle. The sign of the term in brackets depends on the precession angle, which leads to a repulsive interaction ($A > 0$) for $0° \leq \Psi < 54.7°$, and an attractive interaction ($A < 0$) for $54.7° < \Psi \leq 90°$). As discussed in the manuscript, a combination of long-range attraction and a short-range repulsion results in a steady-state distance, $r^*$, where magnetic interaction force between chains equates to zero, $F_m(r^*) = 0$. Consequently, we obtain $\frac{B}{A} = -r^{*k-4}$, which allows us to re-write Eq. S15 as

$$\boldsymbol{F}_{m,ij} = \frac{A}{r^4}\left[1 + \left(\frac{r^*}{r}\right)^{k-4}\right]\hat{\boldsymbol{r}}. \qquad \text{(Eq. S17)}$$

In the simulations, we use the functional form given by Eq. S17, which lets specifying $r^*$ as an input parameter. Eq. S15 tells us that, if $A$ and $r^*$ are known, then the only remaining unknown is $k$, which tunes the stiffness of the repulsive multipolar interaction. We found that setting $k$ to different values in the range of 6 to 8 produce qualitatively similar results in our simulations. Lastly, $\mathbf{v}_{m,ij}$ is calculated by multiplying the magnetic interaction force with the effective mobility of a chain:

$$\boldsymbol{v}_{m,ij} = (6\pi\mu n a)^{-1}\boldsymbol{F}_{m,ij}. \qquad \text{(Eq. S18)}$$

The reduced-order model successfully captures the essential trends observed in experiments: Chains form cohesive clusters with a characteristic steady-state distance between neighbors, clusters performed rotation and translation, and the cluster velocity increased with number of chains. An example of simulated trajectories is displayed in Fig. S6c and Video S7. In order to achieve a quantitative fit between the model and the experiments for translation and angular velocity of clusters, we need to tune two parameters $a_h$ and $h$ (Figs. S6a-b). An experimental determination of $a_h$ and $h$ is difficult due to the anisotropic shape of chains. Also, a simplistic estimation by setting them to half chain length ($L/2 = na$) yields predictions far from the experimental observations. For this reason, we performed simulations for different values of $h$ and re-scaled the simulated values via $a_h$ to match the experimental values. Changing $h$ has a small effect on the translational velocity of clusters (Fig. S6a), and shifts the curve for the angular velocity of clusters by a prefactor approximately proportional to $h^{3/2}$ without changing the shape of the curve (Fig. S6b). The curves in Fig. 5c correspond to $h/L = 0.3$ and $a_h/L \sim 0.26$.

**Supplementary Note 6: Data analysis**

Experimentally, we measure the set of coordinates for the collection of chains $i$, $\{\mathbf{r}_i(t)\}$, at different time points $t$. To find the internal position of chains within the cluster at a desired time point, we need to subtract the mean cluster position and rotation from the set of chain coordinates[7]. Mean cluster position can be obtained by taking the average of chain coordinates, $\mathbf{r}_c(t) = (1/N)\sum_i^N \mathbf{r}_i(t)$. Therefore, chain positions with respect to the moving cluster center can be obtained by, $\mathbf{y}_i(t) = \mathbf{r}_i(t) - \mathbf{r}_c(t)$. To find the optimal mapping between two sets of chain positions measured at consecutive time points, we are required to find the best rotation matrix $\mathcal{R}(t)$ that minimizes the error function, $\sum_i^N [\mathbf{y}_i(t+1) - \mathcal{R}(t)\mathbf{y}_i(t)]^2$, where $t + 1$ denotes the next time point[8]. Finally, internal positions of chains can be obtained after subtracting mean cluster translation and rotation as

$$x_i(t) = \mathcal{R}^T(0) \ldots \mathcal{R}^T(t-2)\mathcal{R}^T(t-1)y_i(t). \tag{Eq. S19}$$

**Rotational order:** Rotational order parameter quantifies the degree of coherence of rotational motion of chains about the cluster center. Rotational order parameter is calculated with the following equation[7],

$$\text{Rotational order} = \frac{1}{N}\left\|\sum_{i=1}^{N}\frac{y_i(t) \times v_i(t)}{|y_i(t) \times v_i(t)|}\right\| \tag{Eq. S20}$$

where $v_i = y(t+1) - y(t)$ is the chain velocity after subtracting the translation velocity of the cluster center. Perfectly coherent rotation results in a rotational order parameter equal to 1, and to 0 for non-coherent motion.

**Connectivity:** Connectivity is calculated based on the idea that strength of cohesive magnetic interactions that holds the chains together would be proportional to the total attractive magnetic dipolar potential in a cluster with the below formula:

$$\text{Connectivity} = \sum_{i \neq j}\frac{n_i n_j}{r_{ij}^3} \tag{Eq. S21}$$

where $r_{ij}$ is the distance between two chains and $n_{i,j}$ is the number of particles in chains $i, j$.

**Mean-squared displacement:** Mean-squared displacement (MSD) is calculated with the standard formula of

$$\text{MSD}(\tau) = \langle \|x_i(t+\tau) - x_i(t)\|^2 \rangle \tag{Eq. S22}$$

where $\langle \cdot \rangle$ is the ensemble average over chains $i$, and time $t$, and $\tau$ is the lag time between two time points. "Fluctuation" quantifies the mean-squared positional fluctuation of chains, *i.e.* deviations around their mean positions in the cluster. Fluctuation is calculated with the following formula[9]:

$$\text{Fluctuation} = \langle \|x_i(t) - \bar{x}_i\|^2 \rangle \tag{Eq. S23}$$

where $\langle \cdot \rangle$ is the ensemble average over chains *i* and time *t*, and $\bar{x}_i$ is the mean position of the $i^{\text{th}}$ chain inside the cluster.

**References**

1. Yung KW, Landecker PB, Villani DD. An Analytic Solution for the Force Between Two Magnetic Dipoles. *Magnetic and Electrical Separation* **9**, 39-52 (1998).
2. Blake JR, Chwang AT. Fundamental singularities of viscous flow. *Journal of Engineering Mathematics* **8**, 23-29 (1974).
3. Swan JW, Brady JF. Simulation of hydrodynamically interacting particles near a no-slip boundary. *Physics of Fluids* **19**, 113306 (2007).
4. Sing CE, Schmid L, Schneider MF, Franke T, Alexander-Katz A. Controlled surface-induced flows from the motion of self-assembled colloidal walkers. *Proceedings of the National Academy of Sciences* **107**, 535-540 (2010).
5. Martinez-Pedrero F, Navarro-Argemí E, Ortiz-Ambriz A, Pagonabarraga I, Tierno P. Emergent hydrodynamic bound states between magnetically powered micropropellers. *Science Advances* **4**, eaap9379 (2018).
6. Tierno P, Schreiber S, Zimmermann W, Fischer TM. Shape Discrimination with Hexapole−Dipole Interactions in Magic Angle Spinning Colloidal Magnetic Resonance. *Journal of the American Chemical Society* **131**, 5366-5367 (2009).
7. Attanasi A, *et al.* Collective Behaviour without Collective Order in Wild Swarms of Midges. *PLOS Computational Biology* **10**, e1003697 (2014).
8. Kabsch W. A solution for the best rotation to relate two sets of vectors. *Acta Crystallographica Section A: Crystal Physics, Diffraction, Theoretical and General Crystallography* **32**, 922-923 (1976).
9. Meinhold L, Smith JC. Fluctuations and Correlations in Crystalline Protein Dynamics: A Simulation Analysis of Staphylococcal Nuclease. *Biophysical Journal* **88**, 2554-2563 (2005).

**Supplementary Figures**

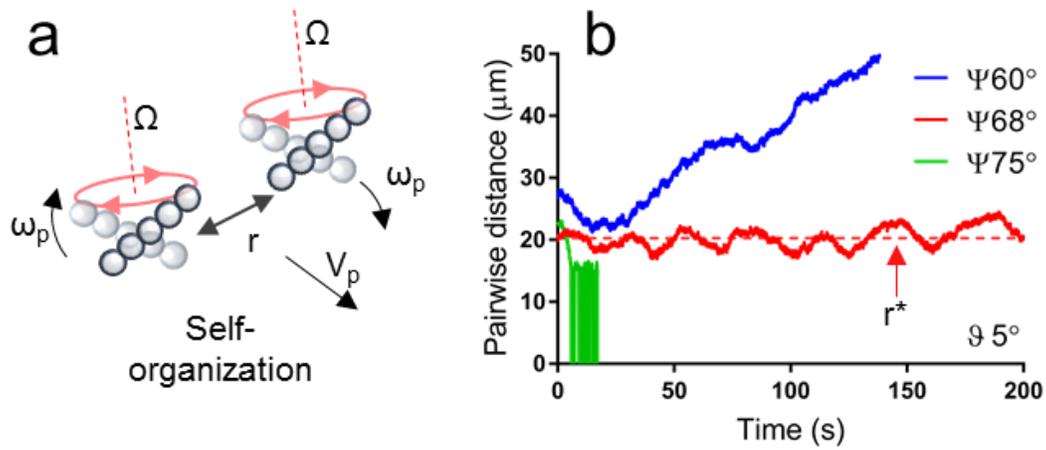

**Figure S1. Experimental characterization of pairwise chain interactions** (a) Schematic of the two-chain system, $r$ is pairwise distance between chains, $V_p$ is pair velocity and $\omega_p$ is pair angular velocity. (b) Pairwise distance measurements under diverging ($\Psi = 60°$), cohesion ($\Psi = 68°$) and collapsing ($\Psi = 75°$) states, where $r^*$ denotes the dynamic steady-state distance.

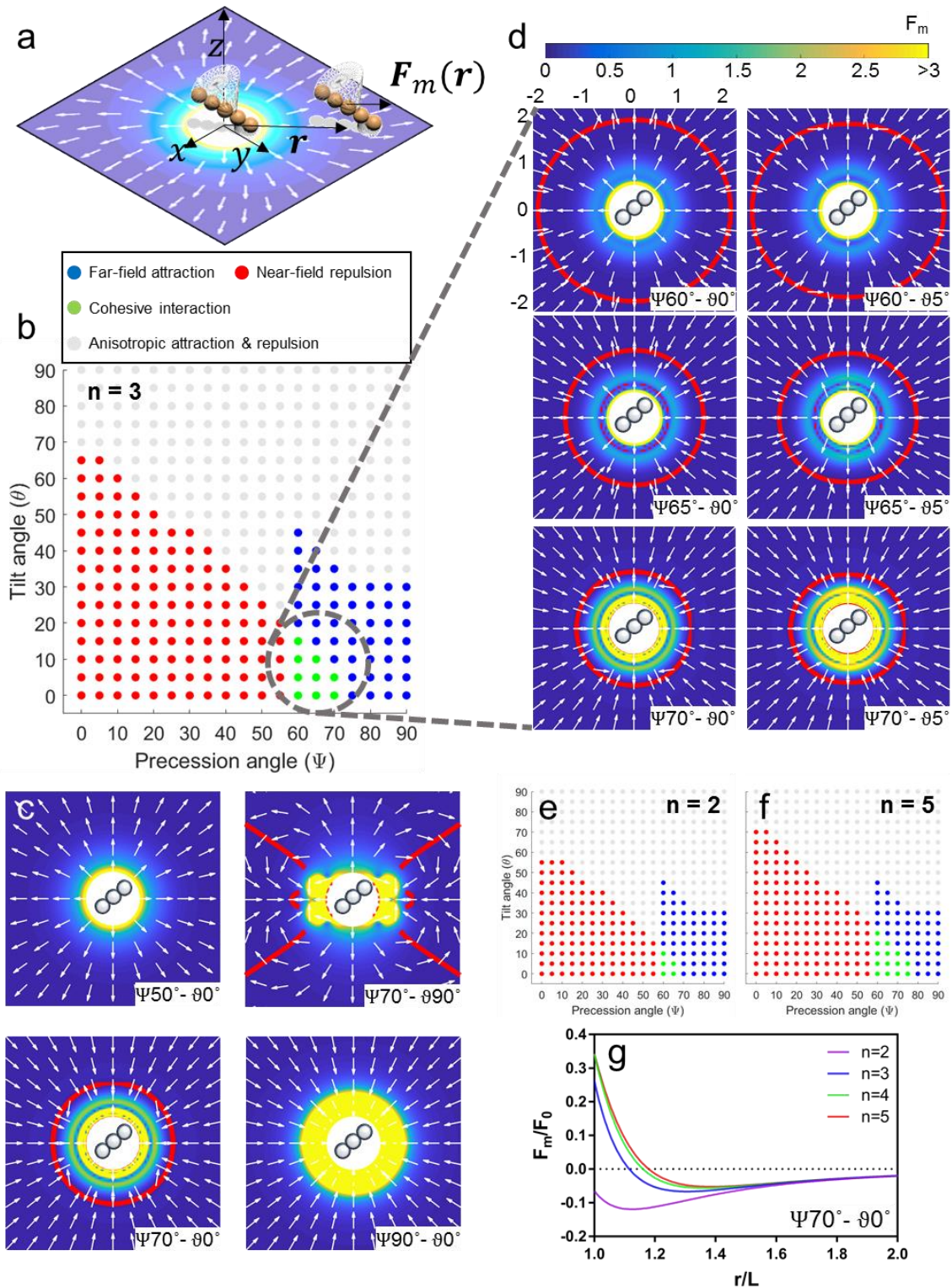

**Figure S2. Pairwise magnetic interactions between chains.** (a) Schematic describes the configuration between two chains used in simulations calculating magnetic interaction forces $F_m$

along the direction of vector $r = (x, y, 0)$ pointing from the first chain at the origin to the second chain. **(b)** Simulated magnetic interaction states between two chains (number of particles per chain, $n = 3$). States are categorized as follows, far-range attraction ($F_m < 0$ for $r/L = 2$), short-range attraction ($F_m > 0$ for $r/L = 1$), cohesive interaction ($F_m < 0$ for $r/L = 2$ and $F_m > 0$ for $r/L = 1$), anisotropic attraction and repulsion ($F_m$ varies between attraction and repulsion at different directions). **(c)** Typical examples from simulations show how the magnetic interactions differ between states. Red curve is where $F_m = 0$, a closed red curve is indicative of cohesive interactions. Arrows indicate the direction of magnetic interaction forces at a given $r$. Color bar indicates the strength of magnetic force, $F_m/F_0$. **(d)** Magnetic interaction force plots for the cohesive states used in the experiments. **(e, f)** Magnetic interaction states for $n = 2$ and $n = 5$ show that the range of $\Psi$ for cohesive interactions change. **(g)** Steady-state distance, $F_m(r^*) = 0$, changes with $n$.

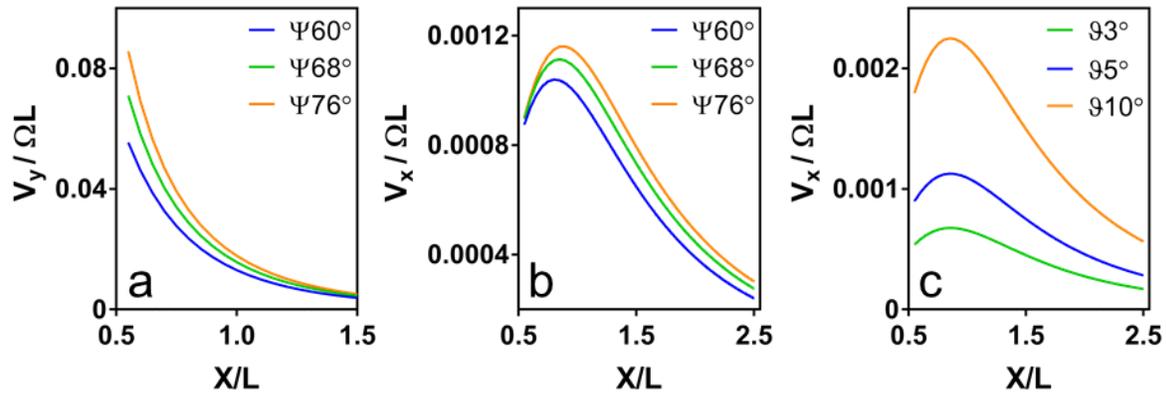

**Figure S3. Flow velocity around a precessing chain near a surface.** Simulated flow velocity is sampled along the line $x > 0$, $y = 0$, $z = 0.5L$ corresponding to the coordinate system shown in Figs. 2c-d. **(a)** Velocity of the rotational flow in $y$ direction resulting from the chain precession about an axis perpendicular to the substrate for different precession angles $\Psi$. **(b, c)** Flow velocity in $x$ direction when the chain precession axis is tilted by angle $\vartheta$. **(b)** $\vartheta = 5°$ for varying $\Psi$, **(c)** $\Psi = 70°$ for varying $\vartheta$.

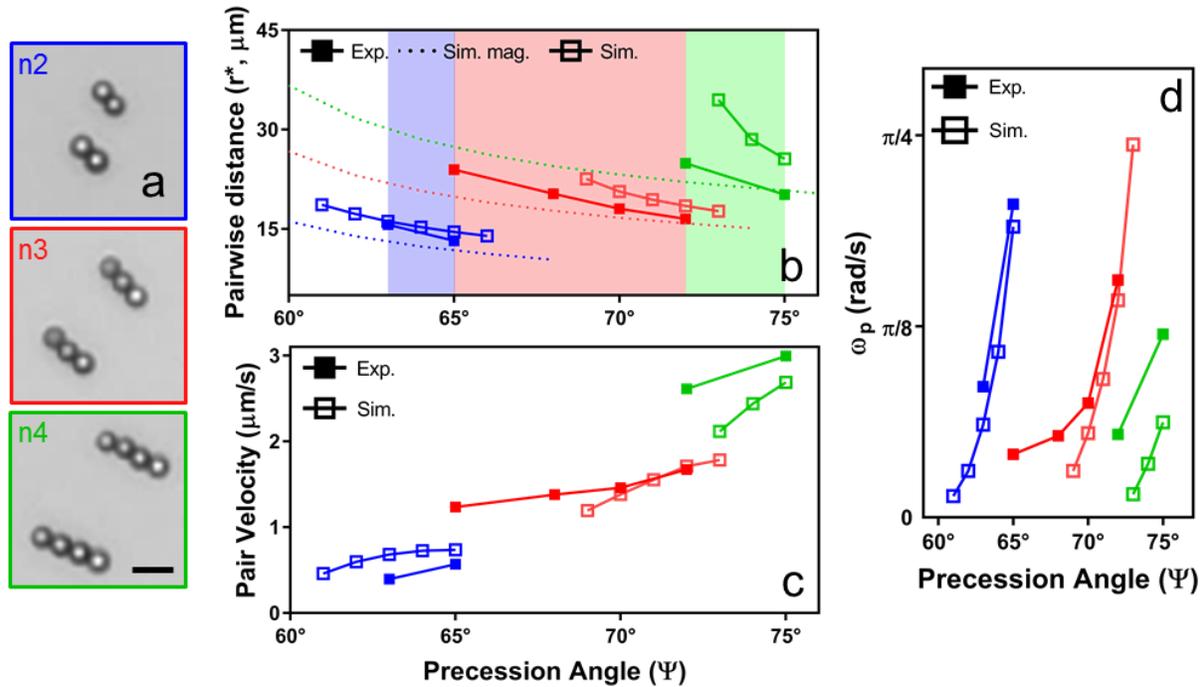

**Figure S4. Experimental and simulated pairwise chain dynamics for different chain lengths.** **(a)** Experimental snapshot of a pair of chains with different number of particles, *n*. Scale bar is 10 μm. **(b)** Pairwise steady-state distance, $r^*$, for different *n* and Ψ. Experimental measurements are compared to simulations including only magnetic interactions (Sim. mag.) and simulations that combined magnetic and hydrodynamic interactions (Sim.). Background color indicates the range of Ψ for experimentally observed cohesive self-organization state (*n* = 2, blue, *n* = 3, red, *n* = 4, green). **(c)** Pair translation velocity and **(d)** pair angular velocity measured from experiments and calculated with simulations. ϑ = 5° for all figures.

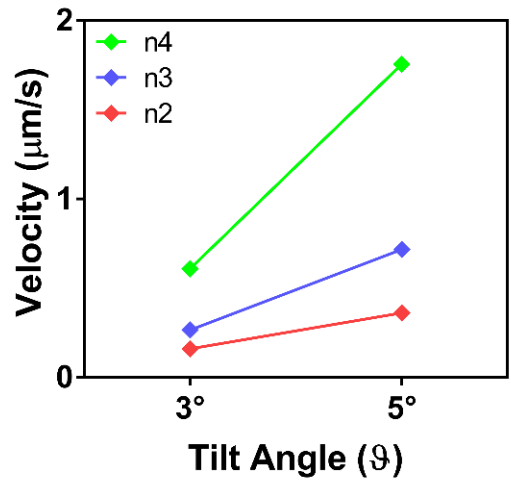

**Figure S5. Experimentally measured single-chain velocities** for different number of beads ($n$) and tilt angles ($\vartheta$) at $\Psi = 68°$.

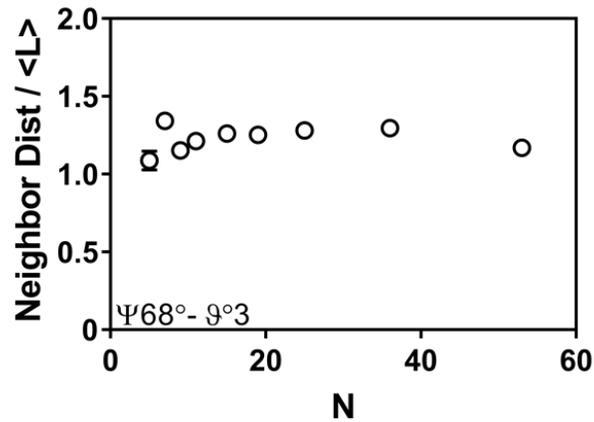

**Figure S6. Mean neighbor distance does not vary significantly with cluster size ($N$).** Nearest neighbors of each chain are detected via Delaunay triangulation. Error bars represent standard deviation of neighbor distances. Distances are normalized to the average chain length, $<L>$.

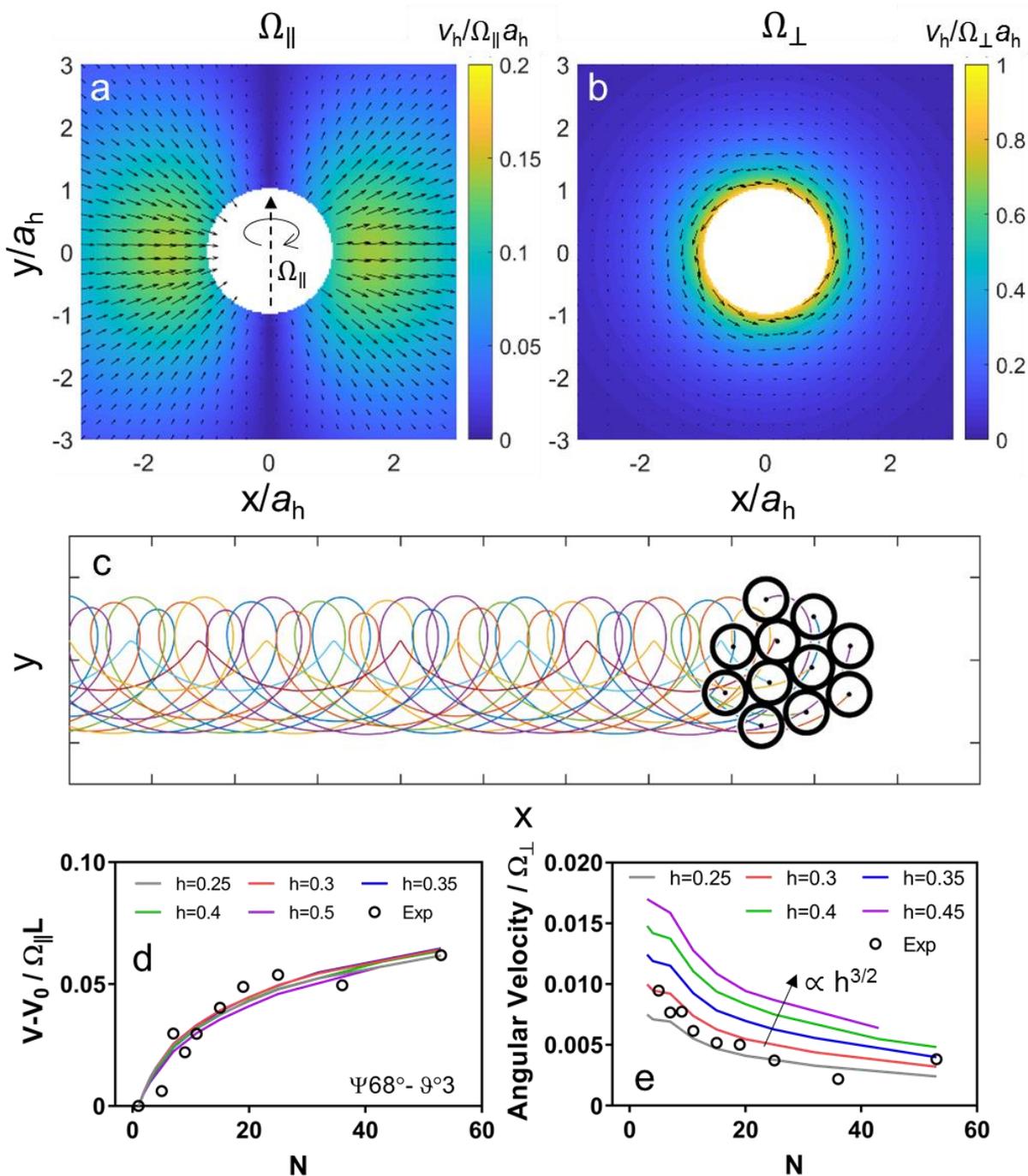

**Figure S7. Reduced-order discrete chain model for simulating the cluster dynamics.** Hydrodynamic field generated by each chain is modeled with a rotlet singularity above a planar no-slip wall. **(a)** Flow generated by a rotlet whose axis of rotation is aligned parallel to the substrate ($\Omega_\parallel$) and **(b)** perpendicular to the substrate ($\Omega_\perp$). Colorbar indicates normalized flow velocity

where $a_h$ is the effective hydrodynamic radius of the chain. **(c)** A typical set of chain trajectories under cohesive interactions, as produced by the reduced order model. Reduced-order model captures experimentally observed changes in **(d)** cluster translation velocity ($V$) and **(e)** angular velocity with increasing cluster size $N$. $v_0$ is the velocity of an individual chain. Model is tuned for different heights of chains from the substrate surface, $h$.

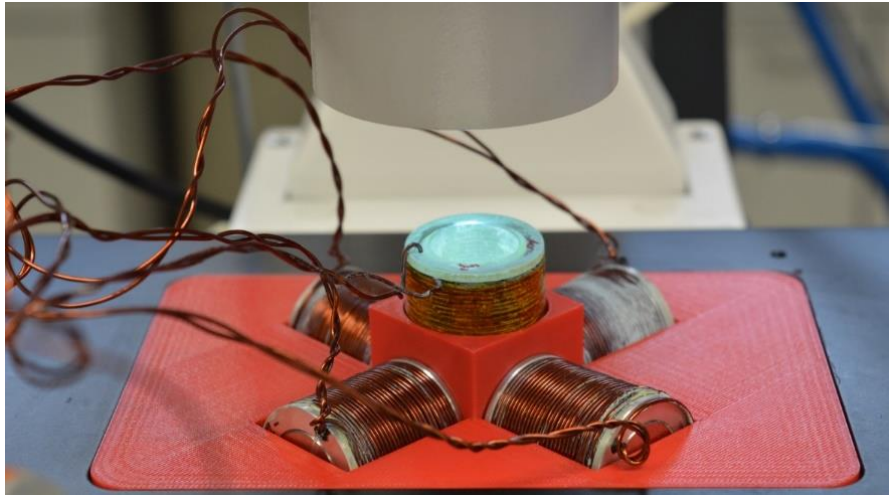

**Figure S8.** Photo of the experimental setup composed of five electromagnetic coils mounted on an inverted optical microscope.